\documentstyle[11pt]{article}

\begin{document}
%

\begin{center}
{\large \bf Ultra High Energy Cosmic Rays in our Universe}
\vskip.5cm

W-Y. Pauchy Hwang\footnote{Email: wyhwang@phys.ntu.edu.tw}$^a$ and Bo-Qiang Ma$^b$ \\
{\em $^a$Asia Pacific Organization for Cosmology and Particle Astrophysics,\\
Center of Theoretical Sciences, Institute of Astrophysics,\\
 and Department of Physics, National Taiwan University, \\
     Taipei 106, Taiwan; \\
$^b$College of Physics, Peking University, Beijing, China}
\vskip.2cm


{\small(24 July 2011; Revised 11 March 2012; 3 November 2016)}
\end{center}

\begin{abstract}

Our Universe is described jointly by Einstein's relativity
principle and the quantum principle; there the existence
of the smallest units of matter, such as electrons,
neutrinos, quarks, and photons, is well established
and the smallest units of matter are described by the
Standard Model. Based on these, we have a clear picture
for the propagation of ultra high energy cosmic rays
(UHECR's), for energies greater than $E > 10^{13}\, eV$
but less than $E < 10^{26}\, eV$, in the cosmic medium
of the Cosmic Microwave Background (CMB) and of the cosmic
background neutrinos (CB$\nu$'s). We find that the CMB plays
a pivot role in this energy range, so that
the observed "knee(s)" and the "ankle" could be
understood in reasonable terms. What we might observe at the
energy near $10^{25}\,eV$', the so-called "$W^\pm$ bursts" or
"$Z^0$ bursts", is briefly discussed. Meanwhile, in the
early Universe, the CB$\nu$'s, in view of the tiny
neutrino mass, gave rise to the clustering phenomenon,
resulting in a neutrino halo attached to each visual
ordinary-matter object, such as the Earth, the Sun,
and stars - with the $PeV$ neutrinos near the surface
of the Fermi-Dirac sphere of the neutrino halo of the
Earth playing the most fundamental role.

\bigskip

{\parindent=0pt PACS Indics: 24.80.+y (Nuclear tests of fundamental interactions
and symmetries); 26.40.+r (Cosmic ray nucleosynthesis); 05.60.-k (Transport
processes); 98.70.Sa (Cosmic rays)}
\end{abstract}

\bigskip

\section{Prelude}

We believe that the cosmic rays, from low-energy muons till ultra
high energy protons and ultra high energy heavy nuclei, are
basically the net results of the interplays of elementary
particles and of heavy nuclei in our Universe. On the other hand,
the $3^\circ\, K$ cosmic microwave background (CMB) is one of
the remnants of the Big Bang that fills, almost uniformly, our
entire Universe, because of the massless of the photon. The
similar story happened to the Cosmic Background (CB) $\nu$'s
(abbreviated for "neutrinos and antineutrinos"), except that
CB$\nu$'s are invisible, or in that it is difficult to see
their manifestations. Basically, they serve the cosmic media
in our Universe. The photons are massless, so are basically
uniformly distributed with currently the temperature close to
$3^\circ\, K$. The neutrinos have a tiny mass and so they
are clustered, in lumps, or in neutrino halos
\cite{Hwang7, Hwang8}.

Given a neutrino halo \cite{Hwang7, Hwang8}, it is a many-body
Fermi-Dirac environment \cite{Kerson} which satisfies Pauli's
exclusion principle. For each neutrino halo, there is a
Fermi-Dirac sphere with some Fermi momentum $k_F^\nu$ -
only neutrinos near the surface of the Fermi-Dirac sphere
can be excited and can interact with the incident
particles. One estimate about the Fermi energy
\cite{Kerson} would be $k_F^\nu\sim 200\,TeV$, a large
energy due to the tiny neutrino mass ($0.058\, eV$)
\cite{Hwang7}.

It is ironic that when we talk about "ultra high energy
cosmic rays (UHECR's)", it is sometime pointing to "the
neutrinos near the surface of the Fermi-Dirac sphere of
a neutrino halo of a visual ordinary-matter object (such
as the Earth, the Sun, and stars)" - a concept introduced
originally in low-energy condensed-matter physics and
in statistical physics \cite{Kerson}. Since these
neutrinos are in fact of the $PeV$ range, they should
indeed be classified "the ultrahigh energy particles".
Thus, we have to think carefully why this Fermi-Dirac
sphere is indeed relevant in the realm of Our Cosmos.

It would be a very important search for $PeV$ neutrinos
or $PeV$ antineutrinos, just like in $IceCube$ \cite{IceCube}.
Granting the existence of the Fermi-Dirac sphere of each
neutrino halo \cite{Hwang7, Hwang8}, this search is
of fundamental importance. Thus, the solar neutrinos
would encounter the antineutrinos of about $PeV$, or
$10^{15}\, eV$, in energy near the surface of the
Fermi-Dirac sphere of the neutrino halo. It would
appear as a spike (i.e., like a very narrow "resonance"),
since other CB$\nu$'s are deep inside the
Fermi-Dirac sphere of the neutrino halo, and hence
are rigid.

What is our Universe? In the 21st Century, we can safely
declare that we are in fact living in the quantum 4-dimensional
Minkowski space-time with the force-fields gauge-group structure
$SU_c(3) \times SU_L(2) \times U(1) \times SU_f(3)$ built-in at
the very beginning. This forms an "overall background". From
this overall background, we can see the lepton world, of atomic
sizes, and we also can see the quark world, of nuclear sizes.
Altogether, we call it as "our Universe", or "our World"
\cite{Hwang417}.

In our Universe \cite{Hwang417}, we have two backgrounds:
the cosmic microwave background (CMB) and the cosmic
background neutrinos (CB$\nu$'s). The $3^\circ \,K$ CMB
are photons, which are massless and are distributed
almost uniformly throughout the Universe (with the
non-uniformity of only one part in $10^5$). On the
other hand, neutrinos (of three flavors) and
antineutrinos have (tiny) masses, so they have
to exist in lumps, or in neutrino halos. They in
fact explain the $25\%$ dark matter \cite{HwangP1},
since neutrinos are the {\it only long-lived}
dark-matter particles in the Standard Model \cite{HwangP1}.

In this beginning of the 21st Century, we have already learnt
to use the $3^\circ\,K$ CMB as a tool to investigate the Universe.
In view of the abundance, sooner or later we should be able to
observe the CB$\nu$'s despite their weak interactions \cite{Hwang7}.
Note that it is the $25\%$ dark-matter neutrinos and antineutrinos
(abbreviated simply as "neutrinos") but only the $5\%$ visual
ordinary matter, quite a contrast.

As we shall see, the high energy electron, or muon, or photon, if higher
than a certain value, will be scattered by the CMB medium and it will not
survive at, e.g., $10^{16}\,eV$. Thus, it would be excluded as a component
of ultra high energy cosmic rays of greater than $10^{16}\,eV$.
Moreover, the ultra high energy cosmic rays might contain things such as
neutrinos and antineutrinos but the reaction $\nu_{UHE} + {\bar \nu}_{free}
\to e^- + e^+$, with the neutrino of a tiny mass such
as $0.058\,eV$, would cut off the UHE neutrinos higher than
$10^{13} eV$, a reasonably low threshold. The possibility
would be $\nu(Solar) + {\bar \nu}(CB; k_F^\nu) \to
e^- + e^+$, capturing the ($PeV$) surface antineutrinos of
the Fermi-Dirac sphere of a specific neutrino halo
\cite{Hwang7, Hwang8}. Finally, there are a few stable
nuclei, including the proton, the alpha particle $^4He$,
the $^{12}C$ particle, etc., those may survive for UHE
energies of $10^{20}eV$ or higher.

In general, we could infer that, at a given high energy, which components
the cosmic rays could have, assuming that the physics is well explained
by nuclear physics and the Standard Model \cite{Hwang417} of particle
physics.

What would be interesting to us in this paper is that our Universe
is not empty, being filled, almost uniformly, by the cosmic medium
of the CMB and filled, non-uniformly, by the CB$\nu$'s in the form
of neutrino halos \cite{Hwang7, Hwang8}. The ultra high energy
cosmic rays, if the energy higher enough, would begin to interact
with the cosmic media of CMB and CB$\nu$'s. The purpose of
this paper is to show that, provided that the Standard Model
\cite{Hwang417} is true, there should be no more mystery in
the forefront of ultrahigh energy cosmic rays.

\bigskip

\section{Ultra High Energy Cosmic Rays}

In our World that is described be the Standard Model
\cite{Hwang417}, the physics of ultra high energy cosmic rays
is controlled by the behaviors of the Cosmic Microwave
Background (CMB) and of the Cosmic Background (CB) $\nu$'s
(abbreviated as "neutrinos and antineutrinos, in three
flavors, and antineutrinos"). Photons are massless and so
the $3^\circ\,K$ CMB is basically uniformly distributed.
Neutrinos have tiny masses and so they followed, in the
early Universe, the visual ordinary-matter heavy lumps
to cluster and to form neutrino halos according to
the Newton's universal gravitational
force \cite{Hwang7, Hwang8}. $PeV$ ($10^{15}\,eV$)
neutrinos would be anticipated near the surface of
the Fermi-Dirac sphere of a neutrino halo attached
to any of the visual ordinary-matter heavy objects
such as the Earth, the Sun, and stars.

The cosmic-rays spectrum at high energies \cite{PDG10, Gaisser} become
quite sophisticated these days, with the so-called "knee", slightly above
$10^{15} eV$, and the "ankle", slightly above $10^{18.5}eV$, appearing
rather convincingly. To develop the field further, it is important to
understand how these phenomena occur, especially in our own Universe.

We are interested in cosmic rays in the energy range greater than
$10^{13} eV$ but less than $10^{26} eV$, including those greater
than $10^{20} eV$ (those unexplored regimes), the so-called
"ultra high energy cosmic rays (UHECR's)".

In particular, we examine the interactions of UHECR with the
$3^\circ\, K$ cosmic microwave background (CMB) or with the clustered
cosmic background $\nu$'s, the cosmic media in our Universe.

Owing to the (possible) neutrino mass of $0.058 \,eV$, the
ultra high neutrino or antineutrino cannot survive if the energy
is greater than $10^{13} \,eV$ (i.e., see the next section or
\cite{Hwang7, Hwang8}), which is rather low in this UHECR game.
On the other hand, electrons (or positrons) would interact with
the CMB photon, without a threshold. The ultrahigh energy
photon could annihilate with a CMB photon, becoming a pair
of $e^+ e^-$ with the threshold of $10^{14}\,eV$ (see
below).

For an UHECR of energy $10^{22}\,eV$ intersecting or
interacting a CMB photon, the center-of-mass (CM)
energy squared would be only $1.3 \times 10^{19}eV^2$.
Similarly, the CM energy would be $1.15 \times 10^{11}\,eV$,
or 115 GeV, if an UHECR of energy $10^{25}\,eV$ intersects
a CMB photon. So, it is slightly above the mass of the $Z^0$
weak boson. The energy range which we talk about
coincides the range which the Standard Model is well
tested \cite{PDG10} - so, we shouldn't anticipate any new
physics but only a replay of the Standard Model physics
in a very peculiar kinematic setup.

The neutrinos have the mass, with the largest about
$0.058\, eV$, much bigger than the temperature of
a few $^\circ K$. Thus, the neutrino mass becomes
a major player in the game. According to the Newton's
gravitational law, an object with mass would have
some universal acceleration near another bigger
object such as the Earth or the Sun.

What is an UHECR particle with the energy
$E\ge 10^{16} \,eV$? According to what we
have said, the list might include the protons,
neutrons (time-dilated), deuterons, alphas,
and heavy nuclei. For some reasons they could
be produced or accelerated to these energies.
As said before, we should exclude, from the list,
electrons (positrons), neutrinos (antineutrinos),
and photons, since the CMB and CB$\nu$'s in our
Universe would cut off the ultrahigh energies.

In the bottom-up scenario, heavy nuclei from
astronomical events, such as from supernova
explosions happening in our direction with high relative
velocities, may provide UHECR's of greater than
$10^{20}\,eV$; chunks of nuclei or protons would
be the origins of those extremely high energy particles
(say, $\ge 10^{22}\, eV$).

The muon, if produced at $10^{22}\,eV$, would have a time-dilation
factor $10^{22}/10^8$, or $10^{14}$; the lifetime would be $2\times 10^{-6}
\times 10^{14} sec$, or $2\times 10^8 sec$ (about 7 years). So, 7 light years
(a muon produced and captured 7 light years away) are still too short in our
astronomical environments. Others such as pions, kaons, etc. have lifetimes
even much shorter and do not play a role here \cite{PDG10}. On the other hand,
a neutron of $10^{22}\,eV$, of which the lifetime at rest is $15\,min$, would
have a dilated lifetime $1,000\, sec \times 10^{22}/10^9$ or $10^{16}\,sec$ or
$3.17\times 10^8$ years. So, neutrons of $10^{22} eV$ would be fairly stable and
could come from 317 Mega light years away or 100 Mpc away.

As said earlier, in our Universe we know that the electrons, positrons,
photons, neutrinos, etc., could not survive beyond certain (high) energies.
In our Universe, there are plenty of $3^\circ K$ cosmic microwave background (CMB) and
clustered cosmic neutrino background (CB$\nu$'s). Even though the energies of
these particles sound extremely low, the ultra high energy cosmic rays (UHECR), including
protons, in fact can see them if energy is high enough.

To obtain some quantitative picture of the various
reactions in the UHECR limits, we proceed to provide
some analysis of these reactions, including the
reactions which do not have the thresholds:

\begin{equation}
e^\pm + \gamma_{CMB} \to e^\pm + \gamma,\qquad \mu^\pm + \gamma_{CMB} \to \mu^\pm
+ \gamma,
\end{equation}

\begin{equation}
p + \gamma_{CMB} \to p + \gamma,
\end{equation}

\begin{equation}
\gamma + \gamma_{CMB} \to \gamma + \gamma,
\end{equation}

\begin{equation}
\alpha + \gamma_{CMB} \to \alpha + \gamma, \qquad etc.
\end{equation}

On the other hand, the reactions listed below have some thresholds and
would start to play some important roles, when UHECR's energy reaches
at the threshold:

\begin{equation}
\gamma + {\gamma}_{CMB} \to e^- + e^+,
\end{equation}

\begin{equation}
p + \gamma_{CMB} \to p + \{ e^- e^+\},
\end{equation}

\begin{equation}
d + \gamma_{CMB} \to p + n,
\end{equation}

\begin{equation}
^3He + \gamma_{CMB} \to d + p,
\end{equation}

\begin{equation}
\alpha + \gamma_{CMB} \to ^3H + p, \qquad etc.,
\end{equation}
plus some others. Hereafter we assume that cosmic rays, depending on the energy, would
be composition of all "stable" particles, including protons, deuterons, $e^\pm$,
$\mu^\pm$, $\gamma$, $\nu$, etc. As said earlier, $\mu^\pm$ may be the borderline of
"stable particles" when we consider the effects due to time dilation; the neutrons,
with the lifetime ($\approx$ 15 min) much longer, could be "stable" if the energy is
greater than, e.g. $10^{22}eV$.

According to our thinking \cite{Hwang7, Hwang8}, there
are plenty of very high energy neutrinos buried in
a Fermi-Dirac sphere of the neutrino halo, associated
with each ordinary-matter heavy object, such as the
Earth, the Sun, and stars. Only those neutrinos near
the surface of the Fermi-Dirac sphere of a neutrino halo,
those $PeV$ neutrinos, may interact with solar neutrinos,
or other particles, and can give rise to some unusual
signals. The detection of $\nu(Solar) +
{\bar \nu}(CB; k_F^\nu) \to e^- + e^+$, using
solar neutrinos \cite{Hwang7, Hwang8} via, e.g., $IceCube$
\cite{IceCube} on the Earth, or on the Venus or on
the Mercury (assuming "zero" experimental backgrounds),
might be so important in verifying the {\it exitence}
of the cosmic background $\nu$'s, by going to a place
with extremely low experimental backgrounds. Assuming
that the Standard Model \cite{Hwang417}, $\nu(Solar)
+ {\bar\nu}(CB; k_F^\nu) \to e^- + e^+$ will be
there. Remember that $\nu(UHE) + {\bar \nu}(free)
\to e^- + e^+$ has the threshold of $10^{13}\,eV$
for the neutrino mass of $0.058\,eV$.

In a related context \cite{Ma}, we discussed the interplay of the cosmic
background neutrinos (CB$\nu$'s) and UHECR's. The significant clustering of
CB$\nu$'s would lead to the {\it first} detection of the CB$\nu$'s. In
\cite{Hwang7, Hwang8}, we realize the {\it existence} of the Fermi-Dirac
sphere of a neutrino halo, attached to the Earth, the Sun, or
stars, so that it should be possible to detect CB$\nu$'s, using
$\nu(Solar) + {\bar \nu}(CB;k_F^\nu) \to e^- + e^+$ provided
that the various low backgrounds could be under control.

\bigskip

\section{Black holes do not exist in our Universe.}

There is an important claim \cite{Hwang7, Hwang8} that
{\it Black holes do not exist in our Universe.} This claim
has something to do the ultra high energy neutrinos (in
three flavors and their antineutrinos) as the surface
neutrinos of the Fermi-Dirac sphere of the neutrino
halo of the Earth (or, of the Venus, or of a star, etc.).
Such surface neutrinos have the Fermi energy as high as
$10^{14-16}\,eV$, in a naive estimate.

Our Universe is dictated by two laws - Einstein's
relativity principle and the quantum principle. At the
beginning of the 21st Century, we realized that there exist
the {\it smallest units of matter}, such as electrons,
neutrinos, quarks, and others, which are described by the
Standard Model \cite{Hwang417}, as first named by S. Weinberg.

In our Universe, there are Cosmic Microwave
Background (CMB) and Cosmic Background Neutrinos
(CB$\nu$'s). Here "neutrinos" stand for neutrinos
of all three flavors and their antineutrinos. CMB
photons, in view of their massless feature, are
almost uniformly distributed, at present nearly
$3^\circ\, K$ with one part in $10^5$ in
fluctuations. Neutrinos have a tiny mass, with
$0.058\,eV$ the largest, so clustering into
neutrino halos.

The world represented by the Standard Model
\cite{Hwang417} is in fact the world of ours.
There is nothing surprising to us - all particles
in there are familiar to us and all interactions are
familiar. That is why we have phrased \cite{Hwang417}:
We declare that we are
living in the quantum 4-dimensional Minkowski space-time
with the force-fields gauge-group structure $SU_c(3)
\times SU_L(2) \times U(1) \times SU_f(3)$ built-in
from the very beginning. This "overall background"
could see the lepton world, of atomic sizes, of the
$SU_L(2)\times U(1) \times SU_f(3)$ symmetry (i.e.,
the other (123) symmetry). It could
also see the quark world, of the nuclear sizes, of
the $SU_c(3) \times SU_L(2) \times U(1)$ symmetry
(i.e., the (123) symmetry).

The other (123) symmetry in the lepton world is required
for the mathematical consistency reason (to eliminate
Landau ghosts and to give the other asymptotic freedom)
as well as for the experimental needs (i.e., neutrino
oscillations and the generation problem, at least).

This is how we describe the {\it smallest} units of
matter. Our world is a world that has the {\it smallest}
units of matter. The language is consistent, and appears
to be complete.

Maybe it is essential to remind ourselves that
a star of five solar-star mass would be an aggregate
of $10^{60}$ smallest units of matter, a really gigantic
large number of the smallest units of matter. Thus, a
possible rescaling, in view of the unknown quantity of
dark matter, of the Newton's gravitational law should
not be a surprise from a theorist's point of view.

Basically, the macroscopic Newton's laws yield
\begin{equation}
m_\nu a = force = G'_N {m_\nu M\over r^2},
\end{equation}
or,
\begin{equation}
a = G'_N {M \over r^2},
\end{equation}
independent of $m_\nu$. So, since the early
Universe, each neutrino halo
would follow the visual ordinary-matter object,
such as a planet or a star, as five times in
weight the invisible dark matter.

Neutrino halos cannot split or fracture by
themselves - since the neutrinos have {\it
only tiny} mass, compared to all the other
particles (except the photon) in the Standard
Model. In the early Universe, they
would follow the formation of the (visual)
ordinary-matter objects - at the end, each
such ordinary-matter object would have one
neutrino halo.

Further evolution of neutrino halos, including
the reactions which we discussed above, appears
only as a very small perturbation. In terms of
many billions of years, the part of neutrino
halos appears to be stable.

As emphasized in an early paper \cite{Hwang7},
the visual ordinary-matter object is an aggregate
of typically $10^{60}$ smallest units of matter,
really macroscopic or gigantic. This factor of
$10^{60}$ is bewildering - so much complexities
at different levels and, yet, so much
simplicities and symmetries.

Thus, CB$\nu$'s are the {\it only long-lived} particles,
provided that the smallest units of matter are
described by the Standard Model \cite{Hwang417}.
It is deduced that, in our World, CB$\nu$'s must be
the $25\%$ dark matter, in the form of neutrino halos.

Neutrino halos are Fermi-Dirac gases, incompressible
because of Pauli's exclusion principle, having the
sizes far bigger than the Schwarzschild size of the
visual ordinary-matter objects (such as stars).
In other words, the point-like structure, such as
the Schwarzschild black hole, would never be formed
in our Universe. The destiny of our World is that
they are filled up with neutrino halos, each
accompanied by a (previously visual) dead star or
a macroscopic object.

So, our World (i.e., our Universe) is basically
to begin with the $3^\circ \,K$ cosmic microwave
background (CMB), with massless photons almost uniformly
distributed coupled with the various neutrino halos, each of
them having a visual ordinary-matter macroscopic object,
such as planets, stars, etc. The incompressible
neutrino halos sort of control the final destiny
of the star systems.

We may safely conclude that black holes
do not exist in our Universe. Neutrino halos have helped to
change the path of the gigantic evolution of the knowledge.

\bigskip

\section{$\nu(Solar) + {\bar \nu}(CB;k_F^\nu) \to e^- + e^+ $}

To verify the existence of the Fermi-Dirac sphere of
a neutrino halo of the Earth (or of the "Venus"), we
have some $IceCube$-like experiment(s) on the Earth
\cite{IceCube}, such that, apart from neutrinos,
there are nothings else - to eliminate all other
experimental backgrounds.

Considering the first-generation "Venus" Satellite
experiments (assuming that the other possible
experimental backgrounds can be vetoed against),
we try to ascertain that there is five times in
weight the dark-matter neutrino halo and such
neutrino halo should have the Fermi-Dirac sphere
with the surface neutrinos of the Fermi energy
$k_F^\nu$ ($\sim 10^{15} \,eV$) detectible
\cite{Hwang7, Hwang8}. This is in fact the other
source of ultra high energy cosmic rays (UHECR's).

On the (real) Venus, we have the following basic information:
\begin{equation}
R=6,051\,Km,\quad M=4.87 \times 10^{24} Kg,\quad T=224.7\, days.
\end{equation}
These numbers are quite close to those for the Earth, the twin
brother ($R=6378\,Km,\,\, M=5.794\times 10^{24} Kg$).

Assuming, for CB$\nu$'s, that the five times of the mass
distributed uniformly over 2.5 times over the radius, we
obtain
\begin{equation}
\rho = 1.68\, g/cm^3 = 5.61 \times 10^{32}\,(eV/c^2)/cm^3.
\end{equation}
There would be a total of $10^{34}$ neutrinos of the
largest mass $0.058\,eV$; divided by six (3 flavors,
plus antiparticles), etc.

The dark matter, 25\% in our Universe as compared
to 5\% of ordinary matter, should be cosmic background
neutrinos and antineutrinos (CB$\nu$'s), according to
the Standard Model \cite{Hwang417} in which CB$\nu$'s
are the {\it only long-lived invisible particles}
\cite{HwangP1}.

We write, for the reaction $\nu + {\bar \nu} \to
e^- + e^+$,
\begin{equation}
\nu(p)+ {\bar \nu}(p') \to e^-(p_e)+e^+(p'_e).
\end{equation}

Using the {\it free} kinematics,
\begin{eqnarray}
k_\mu=({\vec k},i k_0), \qquad & k'_\mu=(0,i m_\nu);\nonumber\\
p_\mu=({{\vec k}\over 2} + {\vec \delta}, i E), \qquad &
p'_\mu=({{\vec k}\over 2} - {\vec \delta}, i E').
\end{eqnarray}
Let's compute the CM energy squared:
\begin{equation}
s \equiv (p_\mu + p'_\mu)^2,
\end{equation}
or,
\begin{equation}
s = (k_\mu + k'_\mu)^2.
\end{equation}
We obtain the threshold energy as follows:
\begin{equation}
k_0 = {1\over m_\nu}(2 m_e^2 + 2 \delta^2 - m_\nu^2),
\end{equation}
indicating that the threshold energy of $10^{13}\, eV$
if $m_\nu = 0.058 \, eV$. Solar neutrinos cannot induce the
reaction since it is far below the threshold energy, if the
the target antineutrino were free. If this would be true,
this would be the first impact of the nonzero neutrino mass.

But the target antineutrinos should be on the
surface of the Fermi-Dirac sphere, of about
$200\, TeV$, well above the threshold
\cite{Hwang7, Hwang8}. If this aspect could
be verified experimentally in some way,
we believe that this would be the fundamental
discovery of physics.

We could use the head-on collision of $\nu(Solar)
+ {\bar \nu}(CB, k_F^\nu)$ as the
sufficient example. That is,
\begin{eqnarray}
k_\mu=( k {\hat z},i k_0), \qquad & k'_\mu=(- k'{\hat z},i k'_0);\nonumber\\
p_\mu=(q {\hat z} + {\vec \delta}, i E), \qquad &
p'_\mu=(q' {\hat z} - {\vec \delta}, i E').
\end{eqnarray}

We could again consider the $s$-squared,
noting that $q$ and $q'$ are much smaller
than $k'_0$ but much larger $k_0$.
\begin{equation}
s \equiv (p_\mu + p'_\mu)^2 \approx - (m_e^2 + \delta^2)
({q'\over q} +2+ {q\over q'}),
\end{equation}
or,
\begin{equation}
s= (k_\mu + k'_\mu)^2 \approx -4 k k'.
\end{equation}
We obtain the threshold energy as follows:
\begin{equation}
4 k k'= (m_e^2 + \delta^2) ({q'\over q} +2+ {q\over q'}),
\end{equation}
remembering that $k'\sim 200\, TeV$ and $k\sim 1 MeV$.

The reaction $\nu_e (Solar) + {\bar \nu}_e \to
e^- + e^+$ takes place because of the $W^\pm$
exchange, or, of the $Z^0$ $s$-pole creation.
It might not be so small (as the ordinary weak
processes) in view of the large energies
involved in the process.

In fact, it is very interesting to work out
the $t$-value:
\begin{equation}
t=(p_\mu-k_\mu)^2 \approx - m_e^2-m_\nu^2 + (m_e^2+\delta^2){k\over q}
+ m_\nu^2 {q\over k}.
\end{equation}
The last term would be the biggest one.

On the other hand, we also have
\begin{equation}
t=(p'_\mu - k'_\mu)^2 \approx -m_e^2-m_\nu^2 +(m_e^2+\delta^2){k'\over q'}
+m_\nu^2 {q'\over k'} + 4 q'k'.
\end{equation}
Here the third and the (last) fifth terms become
the dominant contributions. These are the interesting
plays of the large and small numbers. Note that it seems
to yield the solution that $k<<q'<<q<<k'$.

In the Standard Model \cite{Hwang417}, apart from
the $s$-pole $Z^0$ creation, there is
a cross-generation process, such as $\nu_\mu +
{\bar \nu}_\mu \to e^- +e^+$, via the charged-Higgs
exchange. We assume that this may be smaller than
the $W^\pm$ exchange and the $Z^0$ $s$-pole creation.

Experimentally there is currently the $IceCube$
Collaboration, with a couple of $PeV$ ($1,000 \,TeV$)
detected \cite{IceCube}. Do these neutrinos come from
the surface of the Fermi-Dirac sphere of the
neutrino halo (of the Earth), which we talked about
early on \cite{Hwang7, Hwang8}? The energy range
appears to be correct. Farther vindications appear
to be necessary in this direction.

To some extent, the $IceCube$ "creates" the "Venus"
environment - very low backgrounds for the other
reactions \cite{IceCube}. Of course, we don't know
if the real Venus environment is free of these
experimental backgrounds.

In fact, we should think where many
ultrahigh energy neutrinos would come from. If the
neutrino Fermi-Dirac sphere (of the Earth) does not
exist, the free threshold of about $10\, TeV$
indicated above would cut off the neutrinos
slightly higher than $10\,TeV$ - there should not
be anything observed by $IceCube$ at $1,000\,TeV$.

There are some other channels to detect the Fermi-Dirac
sphere of the neutrino halo of the Earth, or,
of the Venus. For instance, we could consider
$\nu(Solar) + {\bar \nu}(CB; k_F^\nu) \to \gamma
+ \gamma$ against $\nu(Solar) +
{\bar \nu}(free) \to \gamma + \gamma$.

The reaction $\nu(Solar) + {\bar \nu}(CB;k_F^\nu) \to
\gamma + \gamma$ can arise from a few box diagrams with
the charged lepton and $W^\pm$ in the middle. It is
slightly higher order but, based on the Standard Model
\cite{Hwang417}, it would be the leading order for the
annihilation of the $\nu(Solar)-{\bar \nu}(CB)$
pair if the Fermi-Dirac sphere was not there.

We think that the reaction $\nu(Solar) + {\bar \nu}(CB;K_F^\nu)
\to e^- +e^+$ is rather unique - it happens if the Fermi-Dirac
sphere of a neutrino halo exists; it does not occur, otherwise.
However, the need to have a double check is clearly
important.

If we look at the transition amplitude for the reaction
$\nu(Solar) + {\bar\nu}(CB;k_F^\nu)
\to e^- + e^+$, the gauge-boson masses
may not be the only large energy in the problem. Thus, the
transition amplitude, in the U-gauge, should be replaced by

\begin{eqnarray}
& T(\nu_e(Solar) + {\bar \nu}_e(CB;k_F^\nu) \to e^- + e^+)\nonumber\\
=&{G\over \sqrt 2}\{ m_W^2 {\delta_{\lambda\lambda'}+
(q')_\lambda (q')_{\lambda'}/ m_W^2\over (q')^2 +m_W^2-i \epsilon}
i {\bar u}(p_e)\gamma_\lambda(1+\gamma_5)
v(p'_e) +\nonumber\\
& m_Z^2{\delta_{\lambda\lambda'}+q_\lambda q_{\lambda'} / m_Z^2
\over q^2 + m_Z^2 - i\epsilon}
i {\bar u}(p_e)\gamma_\lambda(g'_V + g'_A\lambda_5) v(p'_e)
\}\nonumber\\
& \cdot \sum_j {\bar v}^j(p'_\nu) U^\dagger_{e j}
\gamma_{\lambda'} (1+\gamma_5) \sum_i U_{e i}u^i(p_\nu).
\end{eqnarray}
Here $(q')_\mu = k_\mu-p_\mu$ is the four-momentum
transfer going through the $W^\pm$ boson (in the
$t$-channel) while $q_\mu = k_\mu + k'_\mu$ is
going through the $Z^0$ boson in the $s$-channel.
In the $t-$channel, $(q')^2$ and $q'_\mu$ could
be very large for the ultrahigh energy neutrino:
\begin{eqnarray}
&(q')^2 = ({{\vec k}\over 2}-{\vec \delta})^2 - (k_0-E)^2;
\nonumber\\
& q'_\mu = ({{\vec k} \over 2} - {\vec \delta}, i(k_0-E)).
\end{eqnarray}
In fact, we have, for $k_0 >> m_e >> m_\nu$,
\begin{eqnarray}
& (q')^2  = {k^2\over 4} + \delta^2 - \{\sqrt {k^2 + m_\nu^2}-
\sqrt{(k^2/4)+\delta^2 + m_e^2}\,\}^2\nonumber\\
& = + ( 2\delta^2 + m_e^2 - {m_\nu^2\over 2} + ... ),
\end{eqnarray}
such that, by treating $\delta$ as the same order
as $m_e$, the $(q')^2$ has to be of the same order as
$m_e^2$.

Likewise, the $s-$channel is controlled by
$q^2 = {\vec k}^2 - (k_0^2 + m_\nu)^2
= -(2m_\nu k +m_\nu^2 + ...)$, a competing number
in the range of $\pm m_e^2$ (at $k_0= 10^{13}\,eV$).

The term in $q'_\lambda q'_{\lambda'}/ m^2_W$
or in $q_\lambda q_{\lambda'}/m_Z^2$, when
squared in $\mid T \mid^2$, would appear as
functions of $(q')^2$, or $q^2$, or, perhaps in
interference terms, $q'\cdot q$. As a result,
they are tiny because of the extra
$(q')^2/m_W^2$ or $q^2/m_Z^2$ factor, and
so on.

Similar discussions can be applied to the other
transition amplitude:
\begin{eqnarray}
& T(\nu_\mu(Solar) + {\bar \nu}_\mu(CB;k_F^\nu)
\to e^- + e^+)\nonumber\\
=&{G\over \sqrt 2} m_Z^2 {\delta_{\lambda\lambda'}+
q_\lambda q_{\lambda'}/ m_Z^2\over q^2 +m_Z^2-i \epsilon}
i {\bar u}(p_e)\gamma_\lambda(g'_V+g'_A\gamma_5)
v(p'_e) \cdot \sum_j {\bar v}^j(p'_\nu) U^\dagger_{\mu j}
\gamma_{\lambda'} (1+\gamma_5) \sum_i U_{\mu i}u^i(p_\nu)
\nonumber\\
& + {G'\over \sqrt 2} m_H^2{1\over (q')^2+m_H^2-i\epsilon}
i {\bar u}(p_e)(1-\gamma_5)\times
\sum_i U_{\mu i}u^i(p_\nu) \cdot \sum_j {\bar v}^j(p'_\nu)
U^\dagger_{\mu j}(1-\gamma_5) \times v(p'_e).
\end{eqnarray}
And we have a similar expression for $\nu_\tau(Solar) +
{\bar \nu}_\tau(CB;k_F^\nu) \to e^- + e^+$.

Our discussions lead to the following simplified transition
amplitudes:
\begin{eqnarray}
& T(\nu_e(Solar) + {\bar \nu}_e(CB;k_F^\nu)
\to e^- + e^+)\nonumber\\
=&{G\over \sqrt 2} \{
i {\bar u}(p_e)\gamma_\lambda(1+\gamma_5)
v(p'_e) +\nonumber\\
& i {\bar u}(p_e)\gamma_\lambda(g'_V + g'_A\lambda_5) v(p'_e)
\}\nonumber\\
& \cdot \sum_j {\bar v}^j(p'_\nu) U^\dagger_{e j}
\gamma_{\lambda'} (1+\gamma_5) \sum_i U_{e i}u^i(p_\nu).
\end{eqnarray}

\begin{eqnarray}
& T(\nu_\mu(Solar) + {\bar \nu}_\mu(CB;k_F^\nu)
\to e^- + e^+)\nonumber\\
=&{G\over \sqrt 2}
i {\bar u}(p_e)\gamma_\lambda(g'_V+g'_A\gamma_5)
v(p'_e) \cdot \sum_j {\bar v}^j(p'_\nu) U^\dagger_{\mu j}
\gamma_{\lambda'} (1+\gamma_5) \sum_i U_{\mu i}u^i(p_\nu)
\nonumber\\
& + {G'\over \sqrt 2}
i {\bar u}(p_e)(1-\gamma_5)\times
\sum_i U_{\mu i}u^i(p_\nu) \cdot \sum_j {\bar v}^j(p'_\nu)
U^\dagger_{\mu j}(1-\gamma_5) \times v(p'_e).
\end{eqnarray}

Again, we use the formula for the differential cross section
for the ultrahigh energy neutrinos, say, at $10^{13}\,eV$,
\begin{equation}
{d\sigma\over d\Omega} ={{k^2\over 4}+\delta^2\over (2\pi)^2} \sum_{av}\mid T\mid^2,
\end{equation}
We would obtain
\begin{equation}
\sigma \sim 10^{-31} cm^2,
\end{equation}
a much larger cross section but, for the UHE neutrinos,
this is not subject to observations.

If $PeV$ neutrinos are there, out of the surface of the
Fermi-Dirac sphere of the neutrino halo of the Earth (or,
of the "Venus"), the cross section would be
\begin{equation}
\sigma \sim 10^{-27} cm^2,
\end{equation}
a pleasantly large number.

In the above (Eq. (16)), we emphasize that a neutrino
mass of $0.058\, eV$, the heaviest among the three
flavors, in fact give the threshold of $\sim 10^{13}
\,eV$ on $\nu(UHE) + {\bar \nu}(free) \to e^- + e^+$,
the rather low threshold. This would
place the threshold behavior near $10^{13}\, eV$
for the searches of ultrahigh energy neutrinos
(or antineutrinos) if there is no neutrino halo,
thus no Fermi-Dirac sphere, nearby in the
environment.

As a parenthetical remark, for $\nu_\mu +
{\bar \nu_\mu} \to e^- + e^+$, the
first term, in fact from the $Z^0$ boson, is from the
$s-$channel. The second term, arising the family
Higgs boson, involves the cross-dot products and so
have the $t-$channel behavior.

So far, we have used the U-gauge in our analysis; in the
tree approximations, we note that the results are, in fact,
complete to this order.

In \cite{Hwang7, Hwang8}, we realize that
a neutrino halo, being a Fermi gas, never collapse
and thus the neutrino halo, being five times in weight
as compared to the visual ordinary-matter object,
would stop the black hole from its birth.

\bigskip

\section{Elastic Reactions}

We turn our attention to some analysis of the reactions, from Eq. (1)
to Eq. (4).

Let us quote the Compton-scattering formula \cite{Wu}:

\begin{equation}
\sigma(p+\gamma \to p +\gamma) \to {2\pi\alpha^2\over s} ln({s\over m^2}),
\quad as \quad t \to \infty.
\end{equation}
Now the initial photon is the CMB photon. Suppose that the energy of the UHECR
proton is $10^{22}$ eV and so the CM energy is $s\approx 2\times 10^{22} eV\cdot
6.5 \times 10^{-4}eV \approx 1.3 \times 10^{19} eV^2$. So, we obtain
\begin{equation}
\sigma \approx 10^{-23} eV^{-2} \approx 4 pb \to \lambda_p \approx 2\times 10^5\, Mpc,
\end{equation}
which already exceeds the size of the present Universe (about 4500 Mpc).

On the other hand, $E_p=10^{20}eV$ would imply the mean free path $\lambda\approx
2000\,Mpc$ and $E_p = 10^{18} eV$ implies $\lambda_p \approx 20 Mpc$.

So, for $E\ge 10^{18}\, eV$, the proton is stable, that it does
not get modified by the CMB photons. This is for reaction (2); this is
clearly true for reaction (4), also for heavy nuclei.

Now we turn our attention to the similar formula if the UHECR is the electron
(positron). For the electron energy $E_e=10^{15}\,eV$, we have $s\approx 2\times
10^{15} eV \cdot 6.5\times 10^{-4}\,eV \approx 1.3 \times 10^{12} eV^2$ so that
$\sigma\approx 3.35 \times 10^{-4}\cdot s^{-1} ln (s/m_e^2) \approx 4.14 \times
10^{-16}\,eV^{-2}$, the cross section corresponding to $0.17\, barns$, or to
$\lambda_e \approx 50 kpc$. If nothing else happens, then the electron would be
deflected in $50\,kpc$.

For the muon of energy $10^{19}\,eV$ or $1.3 \times 10^{16} eV^2$, we have
$\sigma\approx 3.35 \times 10^{-4}\cdot s^{-1} ln (s/m_\mu^2)$ or $\sigma\approx
0.7176 \times 10^{-23}\,eV^{-2}$. For such muon, the (dilated) lifetime becomes
$2\times 10^{-6} \times 10^{19-8} sec$ or $2\times 10^5 sec$. Combining the two,
a muon of energy $10^{24}eV$ would last $2\times 10^{10} sec$ and gets negligible
effect from bremsstrahlung.

For the reaction $\gamma + \gamma_{CMB} \to \gamma + \gamma$, it comes
from the box diagrams and is of higher order, $O((\alpha/2\pi)^4)$ (and so is
small). Thus, we needn't consider it for the moment.

Thus, we complete the reactions, Eq. (1) - Eq. (4).

\bigskip

\section{UHECR physics near $10^{20}\, eV$}

Next, we consider those inelastic reactions which might leave their marks
on the UHECR physics, say, Reactions (5)-(9), etc. Reliable estimates can be
obtained by working out the peculiar kinematics and using the well-known cross
sections.

Let us consider, for example, Reaction (6), i.e. $p+ \gamma_{CMB} \to p+
(e^-e^+)$, with $(e^-e^+)$ characterized a composite mass $\bar m$. The
four momentum conservation reads
\begin{equation}
p'_\mu + k'_\mu = p_\mu + k_\mu.
\end{equation}
Or, we have
\begin{eqnarray}
& p'_\| + k'_\| = p+k,\qquad
p'_\bot + k'_\bot = 0,\nonumber\\
& p'+ m_p^2/(2p') + k' + {\bar m}^2/(2k')= p + m_p^2/(2 p) +k.
\end{eqnarray}
We find
\begin{equation}
p'=(4k + m_p^2/p)^{-1} \{ 2p\cdot k + m_p^2- {\bar m}^2/2 \pm [(2p\cdot k)^2
- {\bar m}^2 2p\cdot k + {\bar m}^4/4 -{\bar m}^2 m_p^2]^{1\over 2}\}.
\end{equation}

\begin{equation}
k'=(4k + m_p^2/p)^{-1} \{ 2p\cdot k + m_p^2- {\bar m}^2/2 \mp [(2p\cdot k)^2
- {\bar m}^2 2p\cdot k + {\bar m}^4/4 -{\bar m}^2 m_p^2]^{1\over 2}\}.
\end{equation}

To get some ideas, we have $\sigma(p+\gamma_{CMB}\to p+\gamma) \to {2\pi\alpha^2
\over S} ln ({S\over m^2})$ as $S\to \infty$, with $S\equiv -s$. As quoted
earlier (as our benchmark), at $E\approx 10^{22}eV$, one has $\sigma\approx
10^{-23}eV^{-2}\approx 4 pb$ and, with the density of CMB photons, we find
a mean free path $\lambda\approx 2\times 10^5 Mpc$, bigger than the
Universe size of 4,000 Mpc.

However, the $1\over S$ behavior indicates that at $E\approx 10^{18} eV$ we have
$\sigma \approx 0.4 nb$ or $\lambda \approx 20 Mpc$, a noticeable result. Comparing
the process $p + \gamma_{CMB} \to p + (e^-e^+)$ to $p+ \gamma_{CMB} \to p+\gamma$,
we lose a factor of $\alpha$. This means that at $10^{18} eV$ this effect is barely
visible. {\it This would be a marginal explanation of the ankle effect!!}

Fortunately, there are other reactions, such as $d+ \gamma_{CMB} \to p+ n$
(Reaction (7)) or similar, with the thresholds in the range of a couple of MeV.
The deuteron component in the UHECR flux might be small but the cross section is
much bigger - it serves as an additional reason for the "ankle".

Our explanation of the "ankle" makes some sense. In general,
the electromagnetic effects out of CMB photons, or of higher order, would
make marks in the UHECR physics. On the other hand, the weak reactions, of
cross section $10^{-42} cm^2$ ($= 10^{-6} pb$), are mostly invisible.

Now let us return to Reaction (5), i.e. $\gamma + \gamma_{CMB}\to e^-+ e^+$.
We have
\begin{equation}
S \approx 4 E \cdot E_{CMB} \ge 4 m_e^2; \quad or \quad E_\gamma \ge 4.1 \times
10^{14} eV.
\end{equation}
This means that the high energy photons, those greater than $4.1 \times 10^{14} eV$,
would be depleted from UHECR. After all, the electromagnetic reactions proceed fast
enough. The depletion of the photons from UHECR would explain the happening of the
"knee".

Channel (6) or (7) or others, as described as above in a simplified manner, would
not occur until UHECR reaches a certain threshold. This happens for UHECR at
$10^{18.5}eV$, the
place for the "ankle". In fact, the cross section for the channel $p + \gamma_{CMB}
\to p + (e^-e^+)$ would be down by a factor of $\alpha/\pi$ (as compared to,
for example, Reaction (5) or (7)), but the logarithmic
plot for the UHECR could show that - the effect of $10^{-2.5}$ if protons are
majority of UHECR.

Let come back to Reaction (7), i.e. $d + \gamma_{CMB} \to p + n$. At the threshold,
we find, UHECR identified as deuterons,

\begin{equation}
4 k(p + {m_d^2\over {2p}})+m_d^2 =2 m_n^2 + 2 m_p^2,\quad 4kp =8349.34 MeV^2,\quad
p=8.8785 \times 10^{18} eV.
\end{equation}
These numbers indicate the threshold of $(2m_e)$, or sightly above, to occur at
$10^{18.5}eV$, as explained earlier.

How about reactions (8), (9), etc.? In fact, heavy nuclei ($A\ge 3$), as seen by
the Auger Collaboration \cite{Gaisser, PDG10}, could be of some importance. As
indicated earlier, this may be so if parts of UHECR's come from the inward collapse
of Supernova explosion.

We see that a lot of nuclear reactions with effective energies less than 10 MeV
may become relevant at $10^{19}eV$, until we hit another threshold of the
famous GZK \cite{GZK}:

\begin{equation}
p+\gamma_{CMB}\to \pi + N, \quad E \approx {2m_N m_\pi+m_\pi^2 \over {2E_{CMB}} }
=1.10 \times 10^{20} eV.
\end{equation}

This is another order of magnitude - but very close in our logarithmic plot.
Clearly, interesting physics occurs for UHECR of energy $10^{18.5 - 20.5}eV$.

To say it explicitly, $10^{20}eV$ is where the GZK effect occurs and $10^{18.5}eV$ is
where the "knee" appears (and where nuclear physics dominates). So, what is above
$10^{21-25}\,eV$? Particle physics is probed by CMB - that would be our answer.

The high energy cosmic rays measured in the atmosphere are what we are interested most.
We are already in the vicinity of $10^{20}\, eV$, maybe marching toward higher and
the higher. Those UHECR's may come from the outside solar system, or from the distant
galaxies, and these would be most interesting. As we have said earlier, these UHECR's
are presumably there for a while and thus stable, composed of "stable" particles, such
as $e^\pm$, $\gamma$, $\nu$, $p$, $\bar p$, $d$, ..., $n$, $\mu^\pm$, etc. We don't
take into account $e^\pm$ because of their zigzag paths. we so far don't
take into account $\nu$'s or $\bar\nu$'s mainly due to their (weak) no-interacting
features. As indicated before, too high energy photons (greater than $4.1\times 10^{14}eV$)
could become elusive also. How to reproduce the UHECR curve \cite{Gaisser} should be one of
the most urgent questions.

\bigskip

\section{UHECR Physics near $10^{24}\,eV^2$: $W$-Bursts}

We proceed to consider the reaction $\mu^-(CR) +
{\bar \nu}(CB;k_F^\nu) \to W^*$, where $\mu^-$ comes
from the cosmic rays, maybe of $1\,GeV$, and the CB$\nu$'s
forming the neutrino halo (near the Earth). The Fermi
energy $k_F^\nu$ is assumed to be about $1\,PeV$. So long
as the Fermi-Dirac sphere of a neutrino halo is there, we
should anticipate that the neutrinos (antineutrinos) near
the surface of the Fermi-Dirac sphere could participate
in the various reactions.

This gives another channel of the $UHECR$ physics. In fact,
the muon beam of $1\,GeV$ is readily available at high-energy
accelerators - maybe we could conduct searches on CB$\nu$'s
in terms of the Fermi-Dirac sphere of the neutrino halo of
the Earth.

The reaction $\mu^-(CR) + {\bar\nu}(CB;k_F^\nu) \to
W^*$ should proceed with $s\approx 10^{24}\, eV^2$, assuming
that there exists the neutrino halo of the Earth. So, neutrino
halos (of the Earth, or of the Venus, or of the Sun) opens up
lots of questions.

\bigskip

\section{The Ceiling of the UHECR Physics}

Is there a ceiling on the energy of a particle,
such as the energy of the proton? This is a persistent 
question to ask and to address - we don't see any 
profound reason(s) to say "no" to this question. 

An UHECR particle of energy $10^{25}\,eV$ encountering the $3^\circ\,K$ CMB photon
would have the CM energy squared of $1.3\times 10^{22}\, eV^2$ or the CM energy
$115\, GeV$, just above the $W^\pm$ or $Z^0$ mass. This is where Weiler called
it the $Z^0$-bursts \cite{Weiler}. Clearly, both $W^\pm$ and $Z^0$ show up at
these energies.

If the UHECR particle would be a proton, an alpha particle, or one of those
familiar particles in the Standard Model, it would be a replay of the 
Standard Model \cite{PDG10}, except the very odd kinematics.

What if the UHECR particle is something else, such as some 
supersymmetric particle? But it interacts with the 
$3^\circ\,K$ CMB photon, or with the electromagnetic 
interactions; it means that it carries the ordinary 
electric charge. We infer that this supersymmetric particle
cannot be the lowest-mass neutral supersymmetric particle.
The open-up of such new channel would be fantastic. 
Nevertheless, the Standard Model \cite{Hwang417} is 
working so well and it seems to be so complete and so
self-repulsive, so that the chance of the opening of
another world seems to be rather slim.

\bigskip

\section{Concluding Remarks}

In our opinion, we live in a Universe that is governed by
Einstein's relativity principle and the quantum principle.
It turns out that the smallest units of matter {\it exist}
and are described by these two basic principles. They are
called "the Standard Model of all centuries" \cite{Hwang417},
meaning that the classic thinking of Newton's era will
be replaced by the new framework based on Einstein's
relativity principle and the quantum principle.

The area of Ultra High Energy Cosmic Rays (UHECR's) is one
of the mysterious areas. Since they occur in Our Universe,
they should be described by the Standard Model
\cite{Hwang417}.

Let's emphasize: UHECR's, for the energy
greater than $10^{14} eV$ but less than $10^{26}
eV$, via interactions with the Cosmic Microwave
Background (CMB) and with the Cosmic Background
(CB) $\nu$'s. The CB $\nu$'s provide the best
place for studying the surface of the Fermi-Dirac 
sphere of the neutrino halo of the Earth (or, 
of the Venus) - the fundamental phenomenon which
we should be able to see and to investigate
\cite{Hwang7, Hwang8}. Our Universe is indeed a
peculiar combination of the Bose-Einstein
space-time and the Fermi-Dirac space-time; 
it is just one specific space-time that has 
lot's peculiar properties.

\bigskip

\section*{Acknowledgments}
The Taiwan CosPA project is funded by the Ministry of Education (89-N-FA01-1-0
up to 89-N-FA01-1-5). W-Y. P. Hwang's research is also supported by National Science
Council (NSC99-2112-M-002-009-MY3). B.-Q. Ma's research is supported in part by China's
National Science Foundation.

\end{document}